\pdfoutput = 1
\documentclass[%
reprint,
prd,
amsmath,amssymb,
aps,
superscriptaddress,
floatfix
]{revtex4-1}

\usepackage{dcolumn}
\usepackage{bm}
\usepackage{ulem}
\usepackage{multirow}
\usepackage{placeins}
\usepackage{algorithmic}
\usepackage{color}
\usepackage{comment}

\usepackage{array}
\usepackage{hyperref}
\newcolumntype{P}[1]{>{\centering\arraybackslash}p{#1}}
\usepackage{xcolor}

\begin{document}
\title{Inaccessibility of the Cosmological Singularity to Quantum Probes}
\author{Vishnulal Cheriyodathillathu}\thanks{lal23@iisertvm.ac.in}
\affiliation{School of Physics, Indian Institute of Science Education and Research Thiruvananthapuram, Maruthamala PO, Vithura, Thiruvananthapuram 695551, Kerala, India}
\author{Saurya Das}\thanks{saurya.das@uleth.ca}
\affiliation{Quantum Horizons Alberta \& Theoretical Physics Group, Department of Physics and Astronomy, University of Lethbridge 4401 University Drive, Lethbridge, Alberta T1K 3M4, Canada}
\author{Soumen Basak}\thanks{sbasak@iisertvm.ac.in}
\affiliation{School of Physics, Indian Institute of Science Education and Research Thiruvananthapuram, Maruthamala PO, Vithura, Thiruvananthapuram 695551, Kerala, India}
\author{Tanmay Patil}\thanks{patiltn2601@gmail.com}
\affiliation{Department of Physics, Imperial College London, London}

\begin{abstract}
We propose an operational resolution of the FLRW big bang singularity using quantum probes satisfying a curvature coupled Klein-Gordon equation. Applying the Madelung-Bohm decomposition of a Klein-Gordon field on a classical near-singularity FLRW background, we show that the induced conformal geometry renders curvature invariants finite and makes past directed causal geodesics complete. The classical big bang surface is therefore inaccessible to this class of quantum probes in finite effective proper time, while the underlying FLRW spacetime remains singular.
\end{abstract}

\maketitle

\setlength{\columnsep}{0.5cm}

The standard cosmological model, often called the $\Lambda$CDM model, provides a remarkably successful large scale description of the observable universe. In this framework, the universe is described by Friedmann–Lemaître–Robertson–Walker (FLRW) spacetime which is homogeneous and isotropic and sourced by radiation, baryonic matter, cold dark matter, and a cosmological constant. Precision observations of the Cosmic Microwave Background (CMB), together with large scale structure and distance measurements, are well described by a spatially nearly flat six-parameter $\Lambda$CDM model \cite{Planck:2018bsf}. The discovery of late time cosmic acceleration from Type-Ia supernovae provided the observational foundation for the inclusion of a dark energy component, commonly modeled by $\Lambda$, in the standard cosmological picture \cite{Riess1998,Perlmutter1999}. Despite its observational success, the classical $\Lambda$CDM model inherits one of the deepest conceptual problems of general relativity. When the FLRW equations are extrapolated backwards, the expansion factor vanishes at a finite proper time in the past. At this classical big bang surface, curvature invariants such as the Ricci scalar and the Kretschmann scalar diverge, and causal geodesics are incomplete. This is not merely a coordinate pathology of the FLRW chart and rather it reflects the breakdown of the classical spacetime description. The Hawking-Penrose singularity theorems show that, under suitable causal and energy conditions, geodesic incompleteness is a generic feature of gravitational collapse and cosmology \cite{Hawking:1970zqf}. The Borde-Guth-Vilenkin result sharpens the lesson for expanding cosmologies, since inflation by itself does not remove the need for a past boundary \cite{PhysRevLett.90.151301}. Thus, the initial singularity of FLRW cosmology indicates that classical general relativity cannot provide a complete description of the earliest universe.

Several programs address this objection by changing the dynamics of the geometry \cite{PhysRev.160.1113,PhysRevD.28.2960,PhysRevLett.86.5227,GeorgeFREllis_2004,VILENKIN198225}. However, here we ask a more modest operational question: \textit{Do quantum probes actually experience this classical big bang singularity?} Classical singularities are diagnosed by the behavior of idealized classical point particles and light rays. But actual observations are ultimately mediated by quantum matter. This motivates the possibility that a quantum probe may experience an effective geometry different from the classical metric itself. In the Bohmian-Madelung formulation of relativistic wave dynamics, the phase of a wavefunction defines a momentum field, while the amplitude produces a quantum potential. The resulting quantum-corrected motion can be described as a non-geodesic motion in the original classical background, or equivalently as geodesic motion in a conformally related effective geometry \cite{PhysRev.85.166,Broglie1960NonlinearWM,doi:10.1142/S0217751X98000305,Das:2013oda,Chowdhury:2021fjw}. Recent studies have further explored geometric extensions of Bohmian dynamics \cite{LANTIGUA2025170226,hatifi2025quantumgravitymetricquantization} and operational descriptions of relativistic quantum trajectories based on weak measurements \cite{Foo_2022,PhysRevA.111.052201}.

By a quantum probe we mean a Klein–Gordon test field configuration on the prescribed FLRW background together with the local phase flow defined by its phase. The associated trajectories are integral curves of the phase momentum field. These quantum-corrected trajectories (quantal trajectories) do not, in general, satisfy the geodesic equation of the background spacetime $g_{\mu \nu}$. Instead, they are the geodesics of an effective metric, $\Tilde{g}_{\mu \nu} = \Omega^{2} g_{\mu \nu}$, where the conformal factor $\Omega^{2}$ is determined by the selected field configuration and the specified curvature couplings. In our recent work \cite{Cheriyodathillathu:2026rpz}, this construction was applied to the Schwarzschild black hole singularity. There, the near-core conformal behavior generated by a curvature coupled Klein-Gordon equation was shown to render the effective Kretschmann scalar finite and to push the classical singular surface ($r=0$) to infinite geodesic distance. The resulting resolution is therefore operational and probe dependent. The classical Schwarzschild geometry remains singular, but the corresponding quantum probes do not reach the classical singular surface in finite proper time or affine parameter in the effective spacetime. In the present work, we extend the Bohmian conformal construction to FLRW cosmology and examine whether the classical big bang singularity remains inaccessible to quantum probes.

We begin with a fixed FLRW metric written with the signature (+,-,-,-) as,
\begin{equation}
    ds^{2} = dt^{2} - a^{2}(t) \left[\dfrac{dr^{2}}{1-kr^{2}} + r^{2} d\Omega_{(2)}^{2}\right]~.
\end{equation}
The spatial hypersurfaces of the above spacetime can have positive, zero, or negative curvature for $k=+1,0,-1$ respectively. The prefactor $a(t)$ (expansion factor) determines the overall scale of the spatial metric. We focus on an early classical FLRW branch sourced by a perfect cosmic fluid satisfying the strong energy condition (SEC) and described by an equation of state, $p=\omega \rho$. From the Friedmann equations, such a source gives a decelerating expansion near the big bang surface. Using $\rho\propto a^{-3(1+\omega)}$, the Friedmann equation may be written, up to an overall constant normalization of the density, as \cite{Padmanabhan:2010zzb},
\begin{equation}
    \dot{a}^{2} = \dfrac{8\pi G}{3} a^{-(1+3\omega)} - k ~.
    \label{fried}
\end{equation}
For a spatially flat universe dominated by a single perfect fluid with constant $\omega$, this equation gives the standard power law behavior, $a(t)=a_{0}t^{q}$, where $q=\dfrac{2}{3(1+\omega)}$. Moreover, for $\omega>-1/3$, the density term in Eq.(\ref{fried}) dominates over the curvature term in the limit $a\to0$, the so called big bang singularity. Therefore, the same power law form gives the leading asymptotic behavior near the classical big bang surface even when $k\neq0$. The SEC, together with $\rho>0$, implies $1+3\omega\geq0$, and hence restricts the decelerating singular branch to $0<q\leq1$. 
%The familiar radiation and dust dominated cases correspond to $q=1/2$ and $q=2/3$, respectively. 
We shall use this power law behavior as the local asymptotic description of the classical SEC-satisfying FLRW branch near the singular surface. If an inflationary epoch is included in the subsequent cosmological history, this branch may be interpreted as a possible classical pre-inflationary phase. Similar SEC satisfying pre-inflationary regimes have been considered in several cosmological scenarios \cite{PhysRevD.77.083501,Das:2014ffa,PhysRevD.106.103537}. However, our analysis does not rely on any particular pre-inflationary model. It only uses the local $t\to0^{+}$ behavior of a decelerating classical FLRW branch, hence, an inflationary phase beginning at a finite later time does not affect this near singularity limit. A preferred global positive frequency decomposition need not exist in FLRW spacetime. However, the present construction uses the local phase flow of a selected field configuration and does not require a globally defined positive frequency decomposition.

We now introduce a quantum scalar probe on this fixed spacetime background. The probe is assumed not to back react on the geometry; the scale factor, $a(t)$, is therefore determined entirely by the classical cosmic fluid discussed above. Motivated by the effective-field-theory viewpoint, we allow the probe equation to contain local scalar curvature couplings and consider,
\begin{equation}
     \left(\Box + \dfrac{m^{2}}{\hbar^{2}} + \xi R_{g} + \eta \ell_P^2 K_{g} \right) \Phi = 0~,
     \label{fe}
\end{equation}
where, $R_{g}$ and $K_{g}$ are the Ricci and Kretschmann scalars associated with the classical background spacetime and we take the $\eta>0$ case. Since the background is homogeneous and isotropic, and since our aim is to obtain an effective geometry preserving the FLRW form, we restrict to homogeneous probe states. We therefore write the scalar field in the Madelung-Bohm form,
\begin{equation}
    \Phi(t) = \mathcal{R}(t) e^{i\mathcal{S}(t)}
    \label{anz}
\end{equation}
where $\mathcal{R}(t)$ and $\mathcal{S}(t)$ are real functions of cosmic time. This restriction ensures that the Bohmian conformal factor depends only on time, so that the effective metric experienced by the probe remains within the FLRW class. One can define a phase momentum field $p_{\mu}$, whose integral curves constitute the Bohmian type trajectories considered in this work \cite{Cheriyodathillathu:2026rpz}.
\begin{equation}
    p_{\mu} = \hbar \nabla_{\mu}\mathcal{S},
\end{equation}
\begin{equation}
    U_{\mu} = \dfrac{p_\mu}{\mathcal{M}},  \hspace{30pt} U_{\mu} U^{\mu} = 1~,  
\end{equation}
\begin{equation}
     \mathcal M^{2} = m^{2}+ \hbar^{2} \dfrac{\Box \mathcal{R}}{\mathcal{R}} + \hbar^{2} \xi R_{g}  + \hbar^{2} \eta \ell_P^2 K_{g}
    \label{omega21}~.
\end{equation}
Substituting the ansatz in Eq.(\ref{anz}) to Eq.(\ref{fe}) and separating imaginary and real parts, we obtain,
\begin{equation}
    \dot{\mathcal{S}} = \dfrac{D}{a^{3} \mathcal{R}^{2}}~,
    \label{sdot}
\end{equation}
\begin{equation}
    \ddot{\mathcal{R}} + 3H\dot{\mathcal{R}} - \mathcal{R} (\dot{\mathcal{S}})^{2} + \dfrac{m^{2}}{\hbar^{2}} \mathcal{R}+ \xi R_{g} \mathcal{R} + \eta \ell_P^2 K_{g} \mathcal{R}  =0~,
    \label{real1}
\end{equation}
respectively, where $D$ is an integration constant and $H$ is the Hubble parameter. We are interested in the asymptotic regime $t \to 0^{+}$, corresponding to the classical big bang singular surface. In this limit, for the power law branch $a(t)=a_{0}t^{q}$, the Ricci scalar behaves as $R_{g} \propto t^{-2}$ and Kretschmann scalar behaves as $K_{g}\propto t^{-4}$. Thus the Kretschmann coupling gives the dominant curvature contribution near the singular surface. The leading behavior of Eq.(\ref{real1}) may then be obtained by assuming a power law ansatz for the amplitude $\mathcal{R}(t)$. Solving this resulting equation, we get the final form of the scalar field in the $t \to 0^{+}$ limit as follows,
\begin{equation}
    \Phi(t) \simeq \left(\dfrac{D^{2}}{\eta \hbar G K_{0} a_{0}^{6}}\right)^{\dfrac{1}{4}} t^{(1-\dfrac{3}{2}q)} \,\exp\left[i\left(\mathcal{S}_{0} \pm \dfrac{\sqrt{\eta \ell_P^2 K_{0}}}{t}\right)\right]
    \label{field}
\end{equation}
Eq.(\ref{field}) displays the leading amplitude and phase of the selected asymptotic branch, subleading phase corrections are omitted.
Given the form of the amplitude for the quantum probe, we can calculate the corresponding conformal factor as follows \cite{Cheriyodathillathu:2026rpz},
\begin{equation}
     \Omega^{2} = \frac{\mathcal{M}^{2}}{m^{2}}
    \label{omega2}
\end{equation}
 The effective metric interpretation becomes meaningful when the phase flow curves satisfy the timelike trajectory condition, $\mathcal M^{2}=m^{2}\Omega^{2}>0$. In the homogeneous sector this condition follows directly from the exact field equations. Combining Eq.(\ref{sdot}) and Eq.(\ref{real1}), we obtain,
\begin{equation}
   \mathcal M^{2}
=\hbar^{2}\dot S^{2}
=\frac{\hbar^{2}D^{2}}{a^{6}\mathcal R^{4}}>0
\qquad (D\neq0) 
\end{equation}
We therefore restrict attention to homogeneous solutions with nonzero conserved flux. Consequently, the effective mass remains strictly positive throughout any connected interval on which the background and the solution are regular. For the selected near-singularity branch in Eq.(\ref{field}), $\Omega^{2}(t) \propto t^{-4}$ with a positive leading coefficient for $\eta>0$. The positivity argument above is specific to homogeneous nonzero flux solutions and is not asserted for arbitrary inhomogeneous Klein–Gordon configurations.

As discussed above, Bohmian phase flow curves need not, in general, coincide with geodesics of the background metric $g_{\mu\nu}$. In the admissible timelike sector, their motion admits an equivalent geodesic description in the conformally related effective metric $\widetilde g_{\mu\nu}=\Omega^2g_{\mu\nu}$, where the conformal factor is determined by the field amplitude and the specified curvature couplings, as in Eq. (\ref{omega2}). For the homogeneous configurations considered here, the phase flow curves are comoving and are geodesics of both metrics. However, the two geometries assign different proper time intervals along these coincident paths, allowing the classical singular boundary to lie at finite background proper time but infinite effective proper time. We now analyze the behavior of the effective spacetime near this boundary.

We begin with the form of the Kretschmann scalar in the conformally related spacetime,
\begin{equation}
    \Tilde{K} = 12\!\! \left[\left(\dfrac{\dot{H}+H^{2}+HQ+\dot{Q}}{\Omega^{2}}\right)^{2} \!\!+\! \left(\dfrac{(H+Q)^{2}}{\Omega^{2}}+\dfrac{k}{a^{2} \Omega^{2}}\right)^{2}\right]
\end{equation}
where,
\begin{equation}
    H=\dfrac{\dot{a}}{a} \hspace{25pt}  Q=\dfrac{\dot{\Omega}}{\Omega}~.
\end{equation}
Assuming the asymptotic forms $\Omega^{2} \propto t^{-n}$ and $a(t) \propto t^{q}$, we obtain the following leading behavior for the Kretschmann scalar of the effective spacetime,
\begin{equation}
    \Tilde{K} \propto t^{2n-4}~.
\end{equation}
Similarly, the Ricci scalar of the conformally related metric has the asymptotic behavior,
\begin{equation}
    \tilde{R} \propto t^{n-2} ~.
\end{equation}
Here, $n$ is the positive exponent controlling the near boundary growth. For the quantum scalar probes considered in Eq.(\ref{field}), the conformal factor obtained from the near-singularity Klein-Gordon dynamics behaves as $\Omega^{2}\propto t^{-4}$. Thus, in the notation above, the probe dynamically generates $n=4$, which satisfies the regularity condition $n\geq2$. Consequently, the curvature scalars of the effective geometry experienced by the scalar quantum probe remain finite in the limit $t\to0^{+}$, even though the corresponding invariants of the classical FLRW metric diverge at the big bang surface.

The energy density of the Klein Gordon probe generally increases as the classical singular hypersurface at ($t=0$) is approached, potentially challenging the validity of the fixed background approximation used in this work. Nevertheless, this issue can be systematically controlled through a constant rescaling of the probe field, $\Phi(x)\rightarrow\Phi_{\epsilon}(x)=\epsilon\, \Phi(x)$. Under this transformation, the conformal factor $\Omega^{2}$ remains unchanged, whereas the energy-momentum tensor scales as $T_{\mu \nu} \rightarrow \epsilon^{2}T_{\mu\nu}$. Consequently, within any cutoff region that excludes the singular surface, the probe energy density can be made arbitrarily small by choosing sufficiently small $\epsilon$, while leaving the conformal factor, and hence the effective geometry experienced by the probe, unchanged under this rescaling. This argument does not establish that one fixed nonzero probe amplitude remains negligible uniformly as $t\rightarrow0$, approaching progressively smaller cutoffs may require a sequence of progressively weaker probes \cite{Cheriyodathillathu:2026rpz}.

The regularity of curvature invariants, however, is not by itself sufficient to establish the absence of singular behavior. A stronger, and strictly speaking the only rigorous criterion is geodesic completeness: causal geodesics must not terminate at the would-be singular boundary after a finite value of affine parameter. Although the conformal curvature invariants remain finite as $t \to 0^{+}$, the coordinate system inherited from the classical FLRW metric still places this surface at the edge of the coordinate domain. It is therefore necessary to determine whether (t=0) represents a genuine boundary of the effective spacetime or is instead pushed to infinite geodesic distance in the conformally related geometry. We now show that, for the conformal factors obtained above, both timelike and null geodesics are complete towards $t=0$. We begin with the conformally related metric in the following form,
\begin{equation}
    d\Tilde{s}^{2} = d\Tilde{t}^{2} - \Tilde{a}^{2}(\Tilde{t}) \gamma_{ij}dx^{i}dx^{j}
\end{equation}
where $d\Tilde{t} = \Omega(t) dt$ and  $\Tilde{a}(t)=\Omega(t) a(t)$ are the effective cosmic time and expansion factor in the conformally related metric. The selected homogeneous phase flow corresponds to comoving timelike geodesics. To examine the causal completeness of the associated effective metric, we now consider its general timelike and null geodesics. The time component of their geodesic equation is,
\begin{equation}
    \dfrac{d^{2}\tilde{t}}{d\tilde{\lambda}^{2}} + \Tilde{a} \frac{d\Tilde{a}}{d\Tilde{t}} \gamma_{ij} \dfrac{dx^{i}}{d\Tilde{\lambda}} \dfrac{dx^{j}}{d\Tilde{\lambda}}= 0 
    \label{geoo}
\end{equation}
The normalization of the geodesic tangent in the effective spacetime gives the following identity,
\begin{equation}
    \left(\dfrac{d\Tilde{t}}{d\Tilde{\lambda}}\right)^{2} - \Tilde{a}^{2} \gamma_{ij} \dfrac{dx^{i}}{d\tilde{\lambda}} \dfrac{dx^{j}}{d\Tilde{\lambda}} = \delta~,
    \label{norm}
\end{equation}
where $\delta=1$ for timelike geodesics and $\delta=0$ for null geodesics. Substituting Eq.(\ref{norm}) into Eq.(\ref{geoo}) and solving it further reduces into the following form,
\begin{equation}
    \Delta \Tilde{\lambda} = \int_{0}^{t_{0}} \frac{\Omega}{\sqrt{\delta+ \frac{\vert C \vert}{\Omega^{2} a^{2}}}} dt ~.
    \label{gen}
\end{equation}
In the case of time like geodesics, the integral can be calculated to find the proper time to the singular surface in the $ t \to 0^{+}$ limit as follows,
\begin{equation}
    \Delta \Tilde{\tau} = \int_{0}^{t_{0}} \frac{t^{-\frac{n}{2}}}{\sqrt{1+ \vert C \vert t^{n-2q}}} dt
    \label{time}
\end{equation}
For null geodesics, the affine parameter required to reach the $t=0$ singular surface in the same limit can be derived as,
\begin{equation}
    \Delta \Tilde{\lambda} = \frac{1}{\sqrt{\vert C \vert}} \int_{0}^{t_{0}} t^{q-n} dt
    \label{null}
\end{equation}

To obtain Eqs.(\ref{time}) and (\ref{null}) from Eq.(\ref{gen}), we used only the asymptotic behavior of the conformal factor, $\Omega^{2}\propto t^{-n}$, together with the near singularity power law form of the classical scale factor, $a(t)\propto t^{q}$. The strong energy condition, supplemented by $\rho>0$, enters at this stage only to restrict the classical FLRW branch to $0<q\leq1$. It then follows immediately that the effective proper time of timelike geodesics and the effective affine parameter of null geodesics required to reach the surface $t=0$ diverge whenever $n\geq2$. For the quantum scalar probes considered in Eq.(\ref{field}), the conformal factor obtained from the near singularity Klein Gordon dynamics satisfies precisely this condition. Hence these probes do not reach the classical big bang surface in finite proper time or affine parameter in the conformally related spacetime.

The geometric interpretation is particularly transparent in terms of the effective scale factor, $\Tilde a(t)=\Omega(t)a(t)$. For the branch obtained above, $\Omega(t)$ diverges faster than the classical scale factor $a(t)$ vanishes. Consequently, $\Tilde a(t)\to\infty$  as $t\to0^{+}$. Thus, the classical big bang surface is not experienced by the quantum probe as a collapsing zero volume state. Instead, it is mapped to an infinitely distant asymptotic region of the effective FLRW geometry. The singularity is therefore avoided not through a bounce of the classical scale factor, but through a quantum induced conformal stretching of the near boundary geometry.

In summary, this analysis shows that the FLRW big bang singularity can be resolved in an operational sense once the role of quantum matter in defining the effective spacetime geometry is taken into account. The field obeys the curvature coupled Klein–Gordon equation, while its phase flow admits a geodesic description in a conformally related effective metric. In this effective geometry, curvature invariants such as the Ricci scalar and the Kretschmann scalar remain finite in the classical $a\to0$ limit, and past-directed causal geodesics are complete. This establishes operational inaccessibility within the specified probe model, while the underlying classical FLRW geometry remains singular.

%%%%%%%%%%%%%%%%%%%%%%%%%%%%%%%%%%%%%%%%%%%%%%%%%%%%
\section{Acknowledgment}
%%%%%%%%%%%%%%%%%%%%%%%%%%%%%%%%%%%%%%%%%%%%%%%%%%%%
This work was supported by the Natural Sciences and Engineering Research Council of Canada. Research fellowships from the Council of Scientific and Industrial Research, India, supported V. Cheriyodathillathu.

\nocite{Cheriyodathillathu:2026rpz}
\nocite{Planck:2018bsf}
\nocite{Riess1998}
\nocite{Perlmutter1999}
\nocite{Hawking:1970zqf}
\nocite{PhysRev.85.166}
\nocite{Broglie1960NonlinearWM}
\nocite{doi:10.1142/S0217751X98000305}
\nocite{Das:2013oda}
\nocite{PhysRevD.77.083501}
\nocite{Das:2014ffa}
\nocite{PhysRevD.106.103537}
\nocite{PhysRev.160.1113}
\nocite{PhysRevD.28.2960}
\nocite{PhysRevLett.86.5227}
\nocite{GeorgeFREllis_2004}
\nocite{Chowdhury:2021fjw}
\nocite{VILENKIN198225}
\nocite{Padmanabhan:2010zzb}
\nocite{LANTIGUA2025170226}
\nocite{hatifi2025quantumgravitymetricquantization}
\nocite{Foo_2022}
\nocite{PhysRevA.111.052201}
\bibliographystyle{ieeetr}
\bibliography{bigbang}

\end{document}